%% file: main.tex
\documentclass[runningheads]{llncs}
\usepackage{graphicx}
\usepackage{cite}
\usepackage{hyperref}
\usepackage{amsmath,amssymb,amsfonts}
\usepackage{algorithmicx}
\usepackage{algpseudocode}
\algtext*{EndIf}
\algtext*{EndFor}
\usepackage{graphicx}
\usepackage{textcomp}
\usepackage{xcolor}
\usepackage{algorithm}
\usepackage{multirow}
\usepackage{multicol}
\usepackage{booktabs}
\usepackage{color}
\usepackage{bbding}
\usepackage{array}
\usepackage[inline]{enumitem}
\begin{document}
%
\title{Selecting the Best Sequential Transfer Path for Medical Image Segmentation with Limited Labeled Data}
\titlerunning{Sequential Transfer for Medical Image Segmentation}
%
\author{Jingyun Yang \and
Jingge Wang\and
Guoqing Zhang \and
Yang Li
}
\authorrunning{J. Yang et al.}
\institute{
Shenzhen Key Laboratory of Ubiquitous Data Enabling,\\
Tsinghua Shenzhen International Graduate School,
Tsinghua University \\
Shenzhen 518055, China\\
\email{yangjy20@mails.tsinghua.edu.cn}
}
\maketitle
%
\begin{abstract}
The medical image processing field often encounters the critical issue of scarce annotated data. 
Transfer learning has emerged as a solution, yet how to select an adequate source task and effectively transfer the knowledge to the target task remains challenging.
To address this,
we propose a novel sequential transfer scheme with a task affinity metric tailored for medical images.
Considering the characteristics of medical image segmentation tasks, we analyze the image and label similarity between tasks and compute the task affinity scores, which assess the relatedness among tasks.
Based on this, we select appropriate source tasks and develop an effective sequential transfer strategy by incorporating intermediate source tasks to gradually narrow the domain discrepancy and minimize the transfer cost.
Thereby we identify the best sequential transfer path for the given target task.
Extensive experiments on three MRI medical datasets, FeTS 2022, iSeg-2019, and WMH, demonstrate the efficacy of our method in finding the best source sequence. Compared with directly transferring from a single source task, the sequential transfer results underline a significant improvement in target task performance, achieving an average of $2.58\%$ gain in terms of segmentation Dice score, notably, 6.00\% for FeTS 2022.
Code is available at the git repository: \href{https://github.com/Hiyoochan/SequentialTransfer}{SeqTran}.

\end{abstract}
\input{sections/introduction}

\input{sections/methods}

\input{sections/experiments}

\input{sections/discussion}
\section*{Acknowledgment}

This work is supported in part by the Natural Science Foundation of China (Grant 62371270)  and  Shenzhen Key Laboratory of Ubiquitous Data Enabling (No.ZDSYS20220527171406015).

%
%
%
%

\bibliographystyle{splncs04}
\bibliography{bib}

\end{document}

%% file: sections/introduction.tex
\section{Introduction}
Advances in deep learning have led to rapid developments in medical image processing.
As training from scratch is not a scalable solution in medical image tasks for insufficient annotated data, transfer learning (TL) is a critical technique in training deep neural networks to address the problem \cite{baweja2018towards,wang2021review}.
To ensure robust representational capabilities, this paradigm requires pre-training on adequate source datasets \cite{ma2024segment,kaur2019improving}.
However, the complex nature of clinical medical imaging tasks renders the identification of an appropriate source task for a given target task challenging.
As shown in Fig.~\ref{transfer}, large-scale public medical datasets are often limited in their inclusion of diverse medical images and disease types to comply with various regulatory standards and ethical considerations \cite{litjens2017survey}.
Directly transferring from such source domains can lead to poor model convergence, causing negative transfer \cite{torralba2011unbiased}.
It's worth noting that some datasets closely align with the target dataset in terms of pathological features but fall short in volume, due to the high cost of labeling disease-specific instances.
Existing works \cite{yang2021joint,navon2021auxiliary} have attempted to incorporate these datasets as auxiliary data by training them alongside the target. 
However, such training paradigms can result in inconsistencies in task-specific predictions \cite{yang2023investigating}.
This naturally leads to a need for effectively leveraging such datasets to address the challenges posed by source domain discrepancy \cite{guan2021domain,abnar2021gradual}.
\input{figures/transfer}

In recent studies on TL \cite{yosinski2014transferable,kornblith2019better}, we've observed that training a model on a generalizable source dataset could ensure robust representation capabilities, and learning on a similar source dataset enhances the model's capacity for accurate image reconstruction.
Meanwhile, Baweja et al. \cite{baweja2018towards} showed that sequential knowledge acquisition could indeed facilitate the learning process of target tasks within the realm of continual learning. 
In light of these insights, we propose a sequential transfer learning strategy to help improve the target task performance by optimizing the use of available rare source medical image data. We pre-train on a designated source task and across a spectrum of intermediate source tasks and sequentially transfer to the target.
By leveraging the rich intermediate representations we bridge the gap between source and target domains. Sequential transfer makes better use of these source tasks to improve adaptability, gradually tuning the features to be more aligned with the target task.

To effectively identify related source tasks and a logical transition to the target domain,
we propose a transfer path selection strategy shown in Fig.~\ref{framework}.
The framework includes the relatedness assessment among tasks and the subsequent selection of the sequential transfer path.
First, we propose an analytical task affinity metric for medical image segmentation tasks to quantitatively predict the transfer performance of source tasks. 
Recent studies have incorporated various methods to estimate the knowledge transferability between the source and target tasks for natural images, such as H-score \cite{bao2019information}, OTCE \cite{tan2021otce}, and LEEP \cite{nguyen2020leep}.
However, these metrics are designed for classification and regression tasks instead of semantic segmentation tasks and rely on features extracted by pre-trained models which results in significant computational cost \cite{li2022finding}.
Instead, we propose a simple but effective task affinity metric by calculating the Wasserstein distance of low dimensional image features and structural similarity (SSIM) score of labels to measure medical source task affinity.
To avoid the exhaustive traversal of all path combinations while preserving the effectiveness of sequential transfer, we employ graph clustering in the construction of our source graph.
Once the graph is constructed, given the target task we estimate the transfer cost of paths and, consequently, select the best sequential transfer path toward the target task.
This systematic approach ensures the enhanced performance of specific target tasks while adapting to various medical image segmentation tasks.
Our main contributions are as follows:
\begin{itemize}
    \item We successfully apply a sequential transfer learning pipeline in the medical image processing field to enhance target performance with an average of 2.58\% gain compared to direct transfer methods.
    \item We propose an analytical task affinity metric for medical image segmentation tasks to quantitatively predict the
    transfer performance of source tasks, and thereby identify the most beneficial source tasks.
\end{itemize}
\input{figures/framework}

%% file: figures/transfer.tex
\begin{figure}[t]
 \centering
 \includegraphics[width=\textwidth]{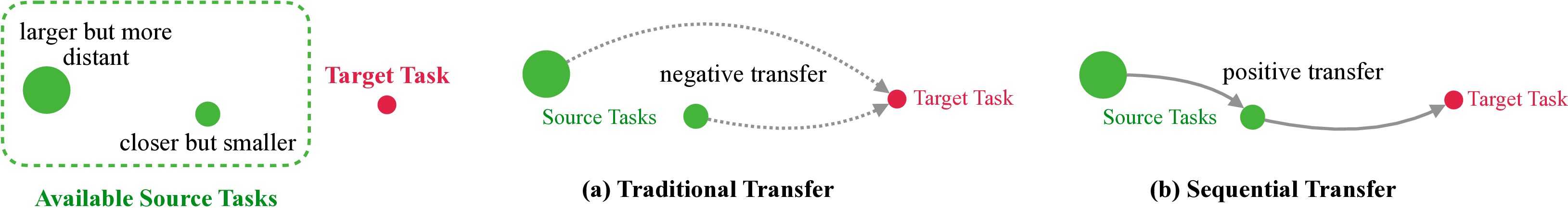}
\caption{Challenges in transfer learning for medical image processing tasks: available source tasks are either large-scale but distant from the target task or closely related to the target task but limited in volume. (a) depicts the traditional direct transfer.  (b) shows the proposed sequential transfer.} \label{transfer}
\end{figure}

%% file: figures/framework.tex
\begin{figure}[t]
 \centering
 \includegraphics[width=\textwidth]{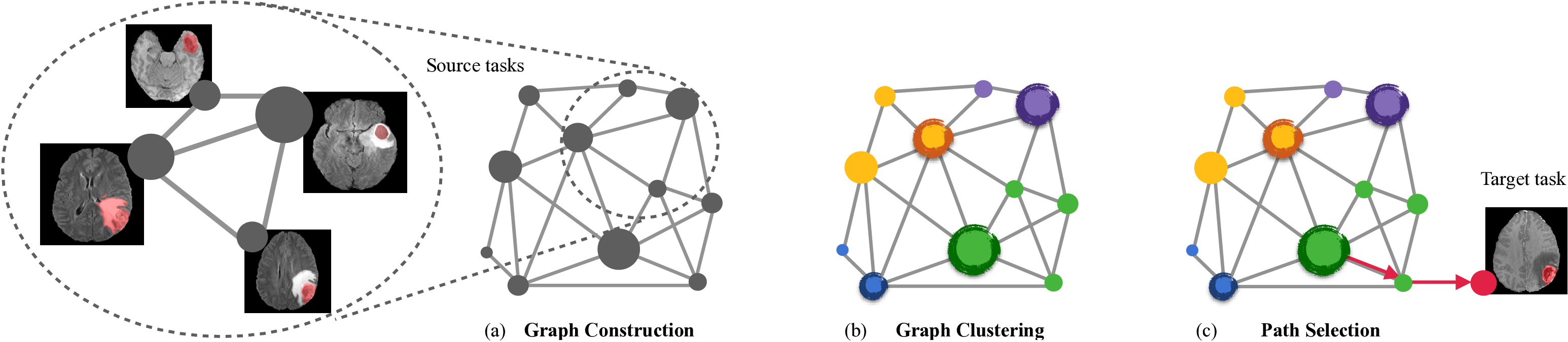}
\caption{Illustraintion of the proposed sequential transfer framework. (a) depicts the graph we construct on source tasks, where edges signify affinity between these medical image processing tasks. (b) groups closely related nodes into clusters. The circled nodes are the representatives of each cluster, which indicate the most generalizable source datasets. (c) illustrates the most effective sequential transfer path we select for a given target task, represented by the bold red line.} \label{framework}
\end{figure} 

%% file: sections/methods.tex
\section{Methodology}

\subsection{Problem Definition}

Suppose we have a set of source tasks $\mathcal{S} = \{s_1,...,s_N\}$ and a target task $t$. For $s_i\in \mathcal{S}$ and $t$, their data are $D_{s_i}=\{(x_j^{s_i},{y}_j^{s_i})\}_{j=1}^{n_i} \sim P_{s_i}(x,y)$ and $D_t=\{(x^t_i,y^t_i)\}^m_{i=1} \sim P_t(x,y)$, respectively.
In the context of TL, given a source model $\theta_s$ we fine-tune its decoder on the target task and obtain transfer accuracy, which can be measured by a segmentation metric (e.g. Dice score), denoted by $\mathcal{A}_{s\to t}$.
The goal is to select an effective sequential transfer path $\mathcal{P}^\star$ for $t$, iteratively fine-tuning the model on each subsequent intermediate source task in the path, to achieve the best transfer accuracy $\mathcal{A}^{\star}$.

\subsection{Task Relatedness Assessment}

First, we construct the graph $G = (V, E)$ on source tasks, where $V\subseteq \mathcal{S}$ represents the set of vertices corresponding to the source tasks, and $E\subseteq V\times V$ is the set of edges, estimated by the task affinity metric.
Given the fact that transferring from the source task with the same modality or same segmentation objective as the target task achieves better results \cite{li2022finding}, we estimate the task affinity with the consideration of both images and labels of medical segmentation tasks. 

\subsubsection{Image Similarity Analysis.}
Given a pair of tasks $(i,j)$ with sample sizes $(N_i,N_j)$, the image characteristics similarity $\mathcal{H}(i,j)$ is measured by the Wasserstein distance \cite{panaretos2019statistical} for its stability and ability to handle shifts in data distributions \cite{yang2023pick}:
\begin{equation}
    \mathcal{H}(i,j) \triangleq  \frac{1}{N_iN_j}\sum^{N_i}_{k=1}\sum^{N_j}_{l=1}\mathcal{W}(\hat{P}_{k},\hat{P}_{l}),
\end{equation}
where $(\hat{P}_{k},\hat{P}_{l})$ are distributions of images $(\hat{x}_k,\hat{x}_l)$ after dimension reduction using principal components analysis.
And the data-pair Wasserstein distance $\mathcal{W}(\hat{P}_{k},\hat{P}_{l})$ is defined as:

\begin{equation}
    \mathcal{W}(\hat{P}_k,\hat{P}_l) = \inf _{\gamma \in \Pi\left(\hat{P}_k, \hat{P}_l\right)} \mathbb{E}_{(x, y) \sim \gamma}\|x-y\|.
\end{equation}
\subsubsection{Label Similarity Analysis.}
We propose to use the structural similarity (SSIM) index \cite{wang2004image} to quantify the similarity of task objectives.
The label similarity $\mathcal{R}(i,j)$ is denoted as:
\begin{equation}
    \mathcal{R}(i,j) \triangleq \frac{1}{N_iN_j}\sum^{N_i}_{k=1}\sum^{N_j}_{l=1}SSIM({y}_k,{y}_l).
\end{equation}
SSIM is often used to evaluate the visual similarity between two images.
Given two voxels $(p,q)$, the data-pair SSIM is:
\begin{equation}
SSIM(p,q)=\frac{\left(2 \mu_{p} \mu_{q}+C_1\right)\left(2 \sigma_{p q}+C_2\right)}{\left(\mu_{p}^2+\mu_{q}^2+C_1\right)\left(\sigma_{p}^2+\sigma_{q}^2+C_2\right)},
\end{equation}
where $\mu_p$ is the average of $p$, $\mu_q$ is the average of $q$,
$\sigma^2_p$ is the variance of $p$, $\sigma^2_q$ is the variance of $q$,
and $\sigma_{pq}$ is the covariance of $p$ and $q$.
$C_1$ and $C_2$ are constants for maintaining stability.
\input{figures/method}

\subsubsection{Task Affinity Estimation.}

For the pair of tasks (i,j), we calculate the task affinity metric $\mathcal{T}_{ij}$ and set the edges accordingly:
\begin{equation}
\label{edge}
    \omega(i,j) = \mathcal{T}_{ij} = \alpha \mathcal{H}(i,j) + \beta \mathcal{R}(i,j),
\end{equation}
where the hyper-parameters set $\{\alpha,\beta\}$ are determined through Bayesian Optimization (BO).
We filter out edges between tasks with neither modality nor segmentation objective in common, as transferring from source tasks with different modalities or limited region of interest (RoI) shape similarity to the target task has been found to be less effective \cite{li2022finding}.
\subsubsection{Graph Clustering.}
After calculating the task affinity scores and setting the edges accordingly, we construct a comprehensive source graph.
To decrease the exhaustive traversal of all path combinations while preserving the effectiveness of sequential transfer, we group closely related source tasks into clusters, where the sequential transfer path is constructed within each cluster, as detailed in the following section.
Viewing all source tasks as nodes, we use the Girvan-Newman algorithm to do graph clustering \cite{girvan2002community} as it is good at detecting hierarchical community structures,
\begin{equation}
    \pi: V \mapsto \{0, 1,..., k-1\}, C_i = \{v_j\in V | \pi(v_j)=i\},
\end{equation}
where the graph clustering algorithm $\pi$ splits the set of nodes $V$ into $K$ clusters $V = \{C_0, C_1,...,C_{k-1}\}$.
Additionally, for TL to work effectively, the source data should be inclusive of sufficient information and have the information accessible.
Thus, we quantify the informativeness of each node $v$ based
on the amount of sample size it holds to ensure sufficient size to capture the complexities and variabilities in the data, denoted by meta-information $\Omega(v)$.
Within each cluster, we set one node to be the representative, accordingly.
The identification of the representative node  $v_{r_i}$ of the cluster $C_i$ is formulated as:
\begin{equation}
    v_{r_i}  = \arg\max_{v_j \in C_i} \Omega(v_j).
 \end{equation}
Following this, we form the representatives set
 $R = \{v_{r_0}, v_{r_1},...,v_{r_{k-1}}\}$. 
As shown in Fig.~\ref{method},
the source graph is pre-constructed, without the need of model training, and prepared for arbitrary target tasks.

\subsection{Sequential Transfer Path Selection}
In the graph context, 
given a target task, we define it as a new node $v_t$.
The goal is to select an effective sequential transfer path $\mathcal{P}$ for $v_t$ with an initial source node $v_0$ from the representatives set $R$ and intermediate source nodes from the same cluster as $v_0$.
For $\mathcal{P}$ whose length is $l$, we formulate the \textit{\textbf{O}ptimal \textbf{S}equential \textbf{T}ransfer} problem as follows:
\begin{equation}
    \begin{aligned}
            OST(D_s,D_t) \triangleq \min_{\substack{v_0\in R;\\v_1,.,v_{l\!-\!1}\in C_{\pi(v_0)}}} 
            \omega(v_0,v_t) + \sum_{i=0}^{l-2}\omega(v_i,v_{i+1}) + \omega(v_{l-1},v_t).
    \end{aligned}
\end{equation}
In this work, we show the sequential transfer result of $l = 2$, as longer paths offer minor improvements while keeping the total training cost manageable.
Then the problem is denoted as:
\begin{equation}
\label{costfunction}
    \begin{aligned}
            OST(D_s,D_t)  \triangleq \min_{\substack{v_r\in R;\\v_i\in C_{\pi(v_r)}}}\omega(v_r,v_t)
              +\omega(v_r,v_i) + \omega(v_i,v_t),
    \end{aligned}
\end{equation}
where we select one initial source node $v_r\in R$ and one intermediate source node $v_i\in C_{\pi(v_r)}$ to find the best sequential transfer path $\mathcal{P}^\star$ with the minimal sequential transfer cost.
When a target task is given, it only requires the computation of edges between the target and source tasks. 
Then we can identify the optimal sequential transfer path using the proposed OST algorithm for the target task.

%% file: figures/method.tex
\begin{figure}[!]
 \centering \includegraphics[width=\textwidth]{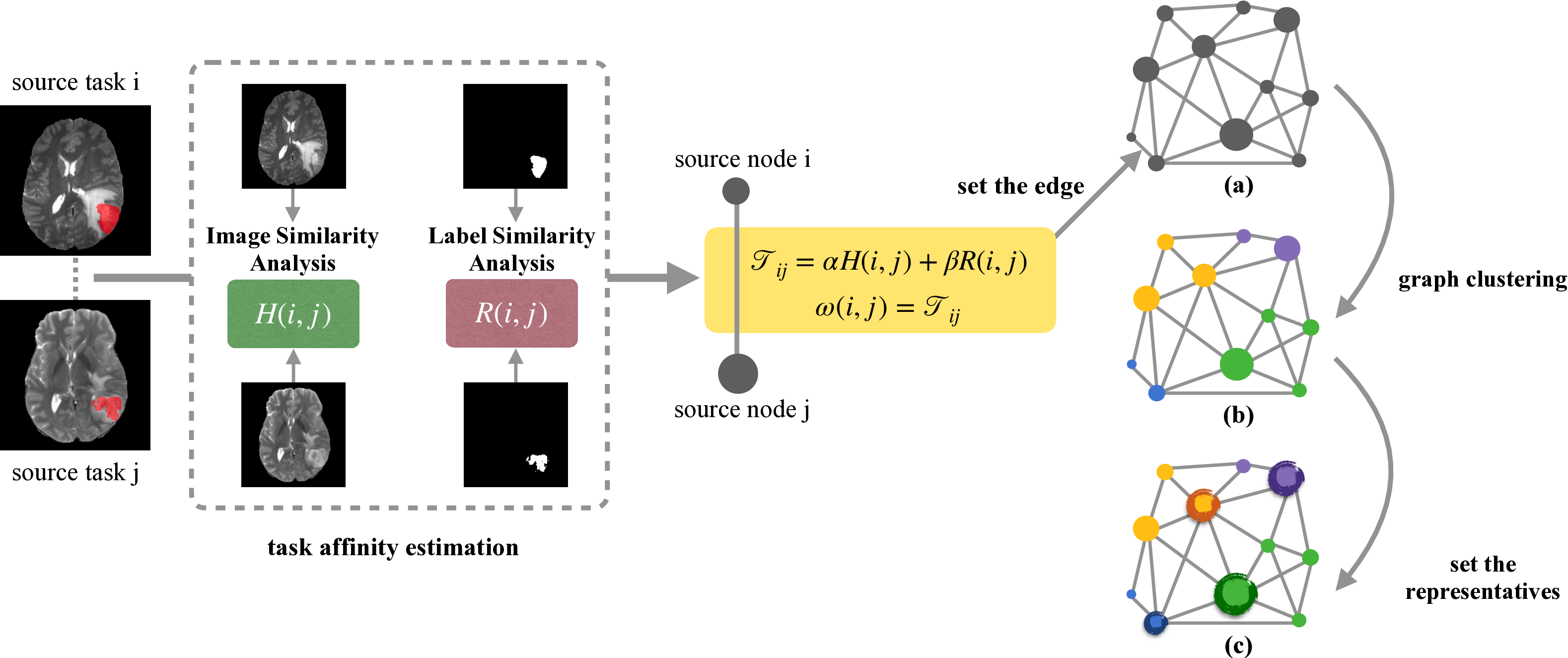}
\caption{Diagram of task relatedness assessment. When a task is represented by a node, the node size indicates the task dataset size while edges signify task affinity.} \label{method}
\end{figure}

%% file: sections/experiments.tex
\section{Experiments and Results}
\subsection{Datasets}
\label{setting}
Three publicly available brain MRI segmentation datasets are used in our work: FeTS 2022 \cite{pati2021federated,baid2021rsna,bakas2017advancing}, iSeg-2019 \cite{sun2021multi}, and WMH \cite{kuijf2019standardized}.
We construct the source graph on FeTS 2022 dataset.
For FeTS 2022, we use MRI volumes across T1, T2, FLAIR, and T1ce modalities, segmenting for enhancing tumor (ET), edema (ED), and necrotic core (NCR), with a resolution of $240 \times 240 \times 155$.
This dataset is split into 22 partitions by the provider, according to different institutions and information extracted from images. Thus, each partition can be seen as an individual domain. 
We select datasets from 8 institutions, each with a sample size exceeding 30, and reorganize the datasets into a collection of binary segmentation tasks on every available modality.
In total, we select 7x4x3 source tasks (from institutions 01, 04, 06, 13, 18, 20, and 21) and 1x4x3 (from institutions 16) target tasks.
Two more datasets are also used as the target datasets.
The iSeg-2019 dataset includes T1, T2 modalities and segments for white matter (WM), gray matter (GM), and cerebrospinal fluid (CSF), with image dimensions of $144 \times 192 \times 256$. 
The White Matter Hyperintensity (WMH) dataset focuses on FLAIR modality for white matter hyperintensities from three institutions, namely, VU Amsterdam (A), NUHS Singapore (S), and UMC Utrecht (U). The volume sizes are $132\times256\times83$, $256\times232\times48$, and $240\times240\times48$ for the three institutes, respectively.
Thus 1x2x3 and 3x1x1 target tasks are also used to perform experiments.
We denote a task as "institute-modality-segmentation objective" and sort the dataset accordingly.
For example, the WMH dataset includes scans from 3 institutes of 1 modality with 1 label, so we sort it into 3x1x1 tasks.

\subsection{Training Setup}
We use the same nnU-Net model architecture \cite{isensee2018nnu} for all experiments and keep the same hyperparameters for different settings.
We follow the prevalent transfer training fashion, which is pre-training the model on a source task and fine-tuning it on the next task. 
During fine-tuning, the encoder is frozen and only the parameters of the decoder are updated.
We pre-train the model on the initial source using 60 samples. 
For fine-tuning, we follow the few-shot training fashion to reduce the size requirements of auxiliary datasets, i.e., intermediate source tasks. 
As each subsequent domain is similar in sequential transfer settings, a few samples yield satisfactory results.  
Therefore, we fine-tune the model with 3 samples for each sequential transfer step, including the final step on the target task. 
\textbf{To ensure} improvements are not solely due to increased training sample size, we maintain the same number of training samples across experiments. 
For instance, in sequential transfer training, we pre-train on 60 samples from the initial source task and fine-tune with 3 samples from the next source task. 
In direct transfer experiments, we use 63 training samples from the source task.
For the target task, we consistently use 3 training samples for all experiments.
Training and test samples are consistently used in compared methods.
Due to the large image size and memory constraints, we set a batch size of 2. 
We crop all data to the region of nonzero values in the same size. 
We use the Adam optimizer with an initial learning rate of $0.01$ and set it to decrease periodically if the losses do not improve enough.
All experiments are conducted on a CentOS 7.6.1810 system with one GeForce RTX 3090 GPU.
\subsection{Performance Evaluation}

We evaluate the proposed sequential transfer framework in comparison with baseline transfer and source model selection methods, organized into the following three parts.
We evaluate the target segmentation performance using the Dice score and IoU. 
The Dice score is computed as $Dice=2TP/(2TP\!+\!FP\!+\!FN)$ and IoU is computed as $IoU={TP}/(TP\!+\!FP\!+\!FN)$.

A quantitative analysis of direct transfer and sequential transfer on three benchmark medical datasets is detailed in Part \textbf{I} of Section.~\ref{seq}, averaging transfer performance on 12 target tasks from FeTS 2022, 6 target tasks from iSeg-2019, and 3 target tasks from WMH.
The experiment settings are:
1) training from scratch on the target task,
2) direct transfer from the initial source task of our OST path,
3) sequential transfer on the selected OST path.
To further investigate the effectiveness of sequential transfer, we detail the transfer performance on each target task in 
Part \textbf{II} of Section.~\ref{when}.
By comparing the best transfer performance of direct transfer with sequential transfer, we aim to determine the scenarios in which sequential transfer proves to be more beneficial.

We also compare different source selection and training methods in transfer learning in Part \textbf{III} of Section.~\ref{path} to show the importance of choosing the appropriate transfer path.
For single-source transfer learning, we directly transfer the knowledge from the selected source task, chosen based on corresponding metrics: 
1) random selection, averaging scores from 10 such choices, 
2) a state-of-the-art transferability metric LEEP \cite{nguyen2020leep},
3) our task affinity metric, selecting the closest source.
For multi-source transfer learning, we learn from multiple source tasks and transfer the knowledge to the target task: 
1) randomly choosing 10 combinations of source tasks and calculating the average score,
2) using a state-of-the-art mix-patch multi-task learning (MTL) model \cite{graham2022one} to jointly train all source tasks from the same OST nodes,
3) sequentially transferring on our selected OST path.
Meanwhile, we also display the ground truth best transfer performance among all source models for both direct transfer and sequential transfer to show if our metric correlates well with the ground truth transfer accuracy.
\input{tables/STL}
\subsubsection{Effectiveness of Sequential Transfer.}
\label{seq}
Table~\ref{STL} shows the segmentation results of different transfer learning methods on three benchmark medical datasets:
FeTS 2022, iSeg-2019, and WMH. The improved performance of sequential transfer on the selected OST path to the target task achieves an average of $2.58\%$ gain in the Dice score, compared to a direct transfer from the initial source task of the OST path.
This is particularly evident for FeTS 2022 with an average gain of 6.00\% in the Dice score.
Take the target task 16-T2-NCR from FeTS 2022 for example, our method achieved a Dice score of 0.5922, surpassing the direct transfer (0.5335) by 11.00\%.
A closer look at the 16-T2-NCR target task offers an illustrative example of the sequential transfer mechanism.
In the previous study \cite{chen2019robust,wang2023learnable}, the T1ce scan is found to be more useful in displaying the enhanced tumor and necrotic core than other modalities.
Interestingly, the OST path chosen for 16-T2-NCR is 01-T1ce-NCR$\to$01-T2-NCR.
This suggests that the model first learns NCR detection well on the T1ce modality.
After that, it strategically shifts to the T2 modality, getting it closer to the target. 
However, direct transfer appears to be more effective for target tasks from the dataset iSeg-2019.
It seems sequential transfer does not always guarantee improved performance over direct transfer.
So to find out when sequential transfer works, we analyze the results of ground truth best performance of both direct transfer and sequential transfer on each target task.

\subsubsection{When does sequential transfer work?}
\label{when}
Sequential transfer does not always guarantee improved performance over direct transfer.
By comparing the best transfer performance of direct transfer with sequential transfer, the results in Tabel~\ref{vs} indicate that sequential transfer is more effective when the imaging modality is not ideally suited for the segmentation task.
We conducted prior experiments on source datasets to determine which modality is better for certain segmentation tasks than other modalities.
The results are shown in Table~\ref{prior}.
\input{tables/prior}
\input{tables/new_dirvsseq}
Take the target task 16-FLAIR-ED for example, since FLAIR is the most suitable modality to perform ED segmentation, direct transfer from the source task 04-FLAIR-ED can adequately acquire the necessary knowledge for feature extraction, as observed from target tasks 16-T1ce-ET, 16-T1ce-NCR, iSeg-T1-WM, and iSeg-T1-CSF, too. 
Nevertheless, the special case is for the target task iSeg-T1-GM, our OST strategy identifies source tasks that both are the T1 modality while the intermediate source node 18-T1-ET is closer to iSeg-T1-GM, enabling the sequential transfer pipeline to work effectively.
Conversely, for other target tasks, the sequential transfer approach is advantageous, enabling the model to sequentially learn features as previously analyzed.


\subsubsection{Effectiveness of OST Path Selection.}
\label{path}
Sequential transfer relies on the accurate selection of source models—a process that, if not executed properly, could lead to suboptimal transfer paths and a decrease in performance gains. 
The results demonstrated in Table~\ref{OST} underscore the importance of careful selection and training of source models.
By comparing different single-source selection methods, we can see that our method could achieve state-of-the-art performance, proving the task affinity metric's capability of accurately estimating the transferability of source tasks.
Moreover, we compare our method with multi-task learning methods to prove that the enhanced performance on the target task is not merely a result of increased training sample size.
Our OST path selection strategy surpasses other methods and correlates well with the ground truth benchmarks for different medical image processing tasks.
For example, our method not only effectively matches the best performance on FeTS 2022, which has the same segmentation tasks as the source, but also identifies effective sequential transfer paths to narrow the domain discrepancy even when target tasks, e.g., iSeg-2019, are quite different from source tasks.
\input{tables/OST}


%% file: tables/STL.tex
\begin{table}[!]
	\centering
\centering
\setlength\tabcolsep{4pt} 
\footnotesize
\caption{Results of different transfer learning training strategies. The fork in Transfer means training from scratch. Dir: direct; Seq: sequential. Bold number: best score.}
\label{STL}
\begin{tabular}{ccccc}
\toprule
\multirow{2}{*}{Target} & \multicolumn{2}{c}{Method} & \multirow{2}{*}{Dice (\%)} & \multirow{2}{*}{IoU (\%)} \\
\cmidrule(lr){2-3}
 & Transfer & \multicolumn{1}{c}{How} & \\
\cmidrule{1-5}
\multirow{3}{*}{FeTS 2022} & \XSolidBrush & - & 63.75 & 59.93 \\ 
 & \Checkmark & Dir & 68.54 & 55.88 \\
 & \Checkmark & Seq & \textbf{72.65} & \textbf{60.72} \\
\cmidrule{1-5}
\multirow{3}{*}{iSeg-2019} & \XSolidBrush & - & 90.27 & 79.33 \\ 
 & \Checkmark & Dir & \textbf{91.66} & \textbf{81.19} \\
 & \Checkmark & Seq & 91.50 & 81.03 \\
\cmidrule{1-5}
\multirow{3}{*}{WMH} & \XSolidBrush & - & 75.71 & 63.68 \\ 
 & \Checkmark & Dir & 79.65 & 68.51 \\
 & \Checkmark & Seq & \textbf{81.19} & \textbf{70.68} \\
\bottomrule
\end{tabular}
\end{table}

%% file: tables/prior.tex
\begin{table}[b]
	\centering
\centering
\setlength\tabcolsep{4pt} 
\footnotesize
\caption{The most ideal modality for each segmentation objective.}
\label{prior}
\begin{tabular}{c|cccc|c|cc|c|c}
\toprule
\multicolumn{1}{c}{\multirow{2}{*}{Task}} & \multicolumn{4}{c}{Modality} & 
\multicolumn{1}{c}{\multirow{2}{*}{Task}} & \multicolumn{2}{c}{Modality} & 
\multicolumn{1}{c}{\multirow{2}{*}{Task}} & \multicolumn{1}{c}{Modality}\\
\cmidrule{2-5}\cmidrule{7-8}\cmidrule{10-10}
\multicolumn{1}{c}{ } & FLAIR & T1 & T1ce & \multicolumn{1}{c}{T2} &
\multicolumn{1}{c}{ } & T1 & \multicolumn{1}{c}{T2} &
\multicolumn{1}{c}{ } & FLAIR\\
\midrule
ED & \Checkmark &   &   &  & WM & \Checkmark &  &   WMH & -\\
ET &   &   & \Checkmark &  & GM & \Checkmark &  & 
 & \\
NCR &  &  & \Checkmark &  & CSF & \Checkmark &  &
 & \\
\bottomrule
\end{tabular}
\end{table}

%% file: tables/new_dirvsseq.tex
\begin{table}[t]
	\centering
\centering
\setlength\tabcolsep{4pt} 
\footnotesize
\caption{Results of best direct and sequential transfer performance on target tasks from three medical datasets: FeTS 2022, iSeg-2019, WMH. Dir: direct transfer; Seq: sequential transfer. Bold number: best score.}
\label{vs}
\resizebox{0.9\columnwidth}{!}{%
\begin{tabular}{lcc|lcc}
\toprule
Target & Method & Dice (\%) & Target & Method & Dice (\%) \\
\cmidrule{1-6}
\multirow{2}{*}{16-Flair-ED}  &  Dir & \textbf{92.46} & 
\multirow{2}{*}{16-T1-ED} &  Dir & 82.08 \\
 &  Seq & 90.57 &
 &  Seq & \textbf{83.75} \\
 \midrule
\multirow{2}{*}{16-T1ce-ED} &  Dir & 84.62 &
\multirow{2}{*}{16-T2-ED} & Dir & 90.49 \\
 &  Seq & \textbf{86.68} &
 &  Seq & \textbf{91.98} \\
 \midrule
\multirow{2}{*}{16-Flair-ET}  &  Dir & 65.65 & 
\multirow{2}{*}{16-T1-ET} &  Dir & 63.23 \\
 &  Seq & \textbf{66.73} &
 &  Seq & \textbf{68.63} \\
 \midrule
\multirow{2}{*}{16-T1ce-ET} &  Dir & \textbf{85.03} &
\multirow{2}{*}{16-T2-ET} &  Dir & 69.27 \\
 &  Seq & 82.18 &
 & Seq & \textbf{73.10} \\
 \midrule
\multirow{2}{*}{16-Flair-NCR} &  Dir & 54.90 &
\multirow{2}{*}{16-T1-NCR} & Dir & 42.15 \\
 &  Seq & \textbf{56.92} &
 &  Seq & \textbf{48.32} \\
 \midrule
\multirow{2}{*}{16-T1ce-NCR} &  Dir & \textbf{80.91} &
\multirow{2}{*}{16-T2-NCR} &  Dir & 53.35 \\
  &  Seq & 75.77 &
  &  Seq & \textbf{59.22} \\
\midrule
\midrule
\multirow{2}{*}{iSeg-T1-WM} & Dir & \textbf{90.83} &
\multirow{2}{*}{iSeg-T2-WM} & Dir & 89.09 \\
  &  Seq & 90.76 &
  &  Seq & \textbf{89.52} \\
\midrule
\multirow{2}{*}{iSeg-T1-GM} &  Dir & 91.48 &
\multirow{2}{*}{iSeg-T2-GM} &  Dir & 90.87 \\
  & Seq & \textbf{92.13} & 
  & Seq & \textbf{91.11} \\
\midrule
\multirow{2}{*}{iSeg-T1-CSF} &  Dir & \textbf{94.71} &
\multirow{2}{*}{iSeg-T2-CSF} &  Dir & 92.80 \\
  &  Seq & 94.45 &
  &  Seq & \textbf{92.92} \\
\midrule
\midrule
\multirow{2}{*}{A-Fl-WMH} & Dir & 73.76 &
\multirow{2}{*}{U-Fl-WMH} &  Dir & 80.12 \\
  &  Seq & \textbf{75.58} &
 & Seq & \textbf{81.70} \\
\midrule
\multirow{2}{*}{S-Fl-WMH} &  Dir & 86.86 & 
\multirow{2}{*}{-} &  - & - \\
  &  Seq & \textbf{87.62} &
 & - & - \\
\bottomrule
\end{tabular}
}
\end{table}

%% file: tables/OST.tex
\begin{table}[!]
	\centering
\centering
\setlength\tabcolsep{4pt} 
\footnotesize
\caption{Dice score (\%) results of different source selection and training methods in transfer learning. Bold number: best score; Red number: best experiment result other than the ground truth best performance.  }
\label{OST}
\resizebox{0.7\columnwidth}{!}{%
\begin{tabular}{cc|ccc}
\toprule
\multicolumn{2}{c}{Method}& \multicolumn{3}{c}{Target} \\
\cmidrule{1-2}\cmidrule{3-5}
Source & How & FeTS 2022 & iSeg-2019 & WMH \\
\midrule
Single & Random & 62.99 & 90.13 & 76.10 \\
Single & LEEP\cite{nguyen2020leep} & 71.50 & 91.32 & 80.10\\
Single & Ours & 70.91 & \color{red}{91.66} & 80.06\\
\midrule
Single & Best & 72.14 & \textbf{91.72} & 80.25 \\
\midrule
\midrule
Multi & Random & 62.10 & 90.21 & 75.54 \\
Multi & MTL\cite{graham2022one} & 71.88 & 91.48 & 79.84\\
Multi & Ours & \color{red}{72.65} & 91.50 & \color{red}{81.19} \\
\midrule
Multi & Best & \textbf{73.65} & \textbf{91.72} & \textbf{81.63} \\

\bottomrule
\end{tabular}
}
\end{table}

%% file: sections/discussion.tex
\section{Conclusion}
In this study, we propose a novel sequential transfer learning method, designed to sequentially transfer from source tasks to a new target task via the best transfer path.
The graph-based sequential transfer path selection strategy identifies the most beneficial source tasks well, thereby ensuring an effective transition sequence in the transfer learning process. Meanwhile, the stepwise learning process of sequential transfer notably improves the target task performance.
The results confirm the efficacy of this approach within the realm of medical image processing.
And we find medical image segmentation tasks benefit from latent sequential learning especially when the target data's modality is not the most suitable for the segmentation objective.  
In the future, we will extend our exploration to include datasets with a broader range of anatomical regions and more medical imaging modalities, such as CT, X-ray, and PET, among others. 